\documentclass[journal,onecolumn]{IEEEtran}
\usepackage{amsmath}                %for great looking maths
\usepackage{amsthm}                 %for great looking theorems, definitions, etc.
\usepackage{amssymb}
\usepackage{thmtools}
\usepackage{kpfonts}
\usepackage[geometry]{ifsym}
\usepackage{dsfont}
\usepackage[mathscr]{euscript}
\usepackage{graphicx}
\usepackage{color}

\usepackage{fancyhdr}
\usepackage{cite}
\usepackage{eurosym}                %For the euro symbol
\usepackage{lscape}                 %landcape pages support --needed for figures!!!
\usepackage{longtable}
\usepackage{booktabs}
\usepackage{color}                  %needed for the hypperref package!
\usepackage[colorlinks=true, citecolor=blue, linkcolor=blue]{hyperref}       %if you want hyperlinks in your pdf document!
\usepackage[font=itshape]{quoting}

\newcommand{\liinesfig}[4]{\renewcommand{\figurename}{Fig.}\begin{figure}[h]\begin{center}\includegraphics[width=#4]{#1}
\vspace{-0.2in}\caption{#2}\label{#3}\vspace{-0.15in}\end{center}\vspace{-0.1in}\end{figure}}
\newcommand{\liinesbigfig}[3]{\begin{figure*}[h!]\begin{center}\includegraphics[width=\textwidth]{#1}
\vspace{-0.25in}\caption{#2}\label{Fig:#3}\end{center}\vspace{-0.25in}\end{figure*}}
\declaretheoremstyle[%
  headfont=\bfseries,%
  headpunct={:},%
  notefont=\normalfont\bfseries,%
  notebraces={--~}{},% punctuation before and after the note
    qed=$\blacksquare$,
]{definitionstyle}
\theoremstyle{definition}
\declaretheorem[style=definitionstyle,name=Definition]{defn}
\declaretheorem[style=definitionstyle,name=Theorem]{thm}

\begin{document}
\title{A Hetero-functional Graph Resilience Analysis for Convergent Systems-of-Systems}
\author{Amro~M.~Farid,~\IEEEmembership{Senior Member,~IEEE,}% <-this % stops a space
%\thanks{Amro M. Farid is with the Engineering Systems and Management Department, Masdar Institute of Science and Technology, PO Box 54224, Abu Dhabi, UAE and the MIT Mechanical Engineering Department
%        {\tt\small afarid@masdar.ac.ae, amfarid@mit.edu}}
}
\markboth{To be submmitted: (Author Preprint) }%
{Retain Empty Field}
\maketitle
\begin{abstract}
Our modern life has grown to depend on many and nearly ubiquitous large complex engineering systems.  Many disciplines now seemingly ask the same question: ``In the face of assumed disruption, to what degree will these systems continue to perform and when will they be able to bounce back to normal operation"?  Furthermore, there is a growing recognition that the greatest societal challenges of the Anthropocene era are intertwined, necessitating a convergent systems-of-systems modeling and analysis framework based upon reconciled ontologies, data, and theoretical methods.  Consequently, this paper develops a methodology for hetero-functional graph resilience analysis and demonstrates it on a convergent system-of-systems.  It uses the Systems Modeling Language, model-based systems engineering and Hetero-Functional Graph Theory (HFGT) to overcome the convergence research challenges when constructing models and measures from multiple disciplines for systems resilience.  The paper includes both the ``survival" as well as ``recovery" components of resilience.  It also strikes a middle ground between two disparate approaches to resilience measurement: structural measurement of formal graphs and detailed behavioral simulation.  This paper also generalizes a previous resilience measure based on HFGT and benefits from recent theoretical and computational developments in HFGT.  To demonstrate the methodological developments, the resilience analysis is conducted on a hypothetical energy-water nexus system of moderate size as a type of system-of-systems.    
\end{abstract}


\IEEEpeerreviewmaketitle

\vspace{-0.1in}
\section{Introduction}\label{Sec:Intro}
\subsection{Motivation:  Need for Resilience in Large Complex Engineering Systems}
Our modern life has grown to depend on many and nearly ubiquitous large complex engineering systems\cite{De-Weck:2011:00}.   Transportation, water distribution, electric power, natural gas, healthcare, manufacturing, agriculture, and supply chain logistics are but a few.  These systems are characterized by a large number of component parts that realize many elaborate processes in a highly complex web of interactions.  Our heavy reliance on these systems coupled with a  growing recognition that disruptions and failures; be they natural or man-made; unintentional or malicious; are inevitable\cite{Hollnagel:2006:00,Leveson:2017:00,Small:2018:00}.   Therefore, in recent years, many disciplines have seemingly come to ask the same question:  ``How \emph{resilient} are these systems?"  Said differently, in the face of assumed disruption, to what degree will these systems continue to perform and when will they be able to bounce back to normal operation\cite{Attoh-Okine:2016:00,Woods:2017:00,Gardoni:2019:00}.  Furthermore, the major disruptions of 9/11, the 2003 Northeastern Blackout, and Hurricane Katrina has caused numerous agencies\cite{Haimes:2008:01,The-White-House-Office-of-the-Press-Secretary:2013:00,Press-Secretary:2011:00} to make resilient engineering systems a policy goal.  

\vspace{-0.1in}
\subsection{Motivation:  Need for Convergent System-of-Systems}
While each of the large complex engineering systems mentioned above require greater resilience in their own right, there is a growing recognition that the greatest societal challenges of the Anthropocene era are indeed intertwined\cite{Wang:2019:10,Farid:2022:ISC-C79,Little:2023:ISC-J52}.  For example, the simultaneous stabilization of carbon emissions, management of nitrogen and phosphorus cycles, and provision of clean water creates complex synergies and trade-offs across multiple underlying engineering systems.   In contrast, a systematic review of 245 publications on the food-energy-water nexus\cite{Albrecht:2018:00} revealed that most do not even capture interactions among water, energy and food that they purport to address.  Similarly, Northeast Hurricane Sandy and Texas Winter Storm Yuri revealed the interdependence of the natural gas, electric power, and transportation systems\cite{Wang:2014:02,Kemabonta:2021:00,Nejat:2022:00}.  The interdependence and consequent emergent behavior between these critical engineering systems must be systematically identified, understood, and analyzed\cite{Rinaldi:2001:00,Rainey:2022:00} within a larger \emph{systems-of-systems} modeling and analysis framework\cite{Little:2019:00,DeLaurentis:2022:00}.  Because of each of these systems is associated with its own engineering discipline, systems-of-systems engineering \cite{Jamshidi:2011:00,Jamshidi:2017:00} requires a deep \emph{convergence} of disciplines founded on reconciled ontologies, data, and theoretical methods\cite{Guizzardi:2005:00}.   

\vspace{-0.1in}
\subsection{Motivation:  Need for Resilience Measures}
Perhaps in contrast to convergence research, a large body of academic literature has developed on the subject across multiple disciplines.  These include ecological\cite{Holling:1973:00}, economic\cite{Rose:2007:00}, organizational\cite{Vogus:2008:00,Naderpajouh:2018:00}, network\cite{Najjar:1990:00,Yodo:2016:01,Cai:2018:00,Abimbola:2019:00,Kammouh:2020:00,Cai:2021:00}, socio-ecological\cite{Diamond:2011:00},  infrastructure\cite{Park:2013:00,Sansavini:2017:00,Hickford:2018:00} , and psychological\cite{VanBreda:2001:00} resilience.  Not surprisingly, a number of reviews \cite{VanBreda:2001:00,Bhamra:2011:00,Righi:2015:00,Hosseini:2016:01,Yodo:2016:02,Patriarca:2018:00,Naghshbandi:2020:00,Penaloza:2020:00,Wied:2020:00} on the topic have found that these contributions while complementary are not necessarily in agreement\cite{Dekker:2017:00}.  The emerging field of resilience engineering, therefore, is still developing and requires formal conceptualizations and definitions\cite{Madni:2009:00,Madni:2011:00,Jackson:2013:00,Woods:2015:00,Hollnagel:2016:00,Cottam:2019:00,Yu:2020:00}, and quantitative models and measures\cite{Youn:2011:00,Ayyub:2013:00,Francis:2014:00,Yodo:2016:03,Yodo:2017:01,Ren:2017:00,Wu:2021:00}.  A key element to such rigorous approaches is the development of resilience measures which many, even recently, have identified as an area for concerted effort\cite{Ayyub:2013:00,Madni:2009:00,Henry:2012:00,VanBreda:2001:00,Whitson:2009:00,Bhamra:2011:00,Barker:2013:00,Francis:2014:00,Pant:2014:00,Reed:2009:00,Madni:2020:00}.  Such resilience measures would not only quantify resilience but could also inform designers and planners in advance how to best improve system resilience\cite{Salomon:2020:00}.  

\vspace{-0.1in}
\subsection{Original Contribution}
This paper develops a methodology for hetero-functional graph resilience analysis and demonstrates it on a convergent system-of-systems.  In that regards, it seeks to address the recognized need for resilience in systems-of-systems\cite{Joannou:2019:00} as a specialized class of engineering systems.  It also seeks to use the Systems Modeling Language\cite{Delligatti:2014:00,Friedenthal:2014:00}, model-based systems engineering\cite{Pyster:2022:00} and Hetero-Functional Graph Theory (HFGT)\cite{Schoonenberg:2019:ISC-BK04,Farid:2022:ISC-J51,Farid:2016:ISC-BC06} to overcome the convergence research challenges when constructing models and measures for systems resilience\cite{Madni:2020:00}.  In addition to addressing this two-fold need, this work is in agreement with the developing majority view which divides the resilience property into two complementary aspects:  a static ``survival" property which measures the degree of performance after a disruption, and a dynamic ``recovery" property which measures how quickly the performance returns to normal operation (Figure \ref{Fig:resilfiig})\cite{VanBreda:2001:00,Bhamra:2011:00,Righi:2015:00,Hosseini:2016:01,Yodo:2016:02,Patriarca:2018:00,Naghshbandi:2020:00,Penaloza:2020:00,Wied:2020:00}.  This paper also strikes a middle ground between two complementary approaches to resilience measurement.  In the first, many resilience measures depend on formal graph theoretic approaches.  The choice of HFGT over formal graph theory allows the paper's scope to expand from \emph{homo-functional} engineering systems to \emph{hetero-functional} engineering systems, and systems-of-systems more specifically.  In the second, resilience measurement is often conducted via complex simulation packages.  While HFGT facilitates a variety of engineering system simulation approaches, the model presented in this work is at a higher level of abstraction to facilitate decision-making earlier in the system life-cycle.  This paper also generalizes a previous resilience measure based on HFGT\cite{Farid:2015:ISC-J19,Farid:2014:ISC-C37,Farid:2014:ISC-C38,Thompson:2020:SPG-C68,Thompson:2021:SPG-J46} which demonstrated its \emph{convergence potential} with applications in transportation, electric power, water distribution, and production systems.  
The resilience analysis presented here also benefits from recent theoretical and computational developments in HFGT\cite{Schoonenberg:2019:ISC-BK04,Farid:2022:ISC-J51,Farid:2016:ISC-BC06,Thompson:2022:ISC-C80,Schoonenberg:2022:ISC-J50,Thompson:2023:ISC-JR02} that directly support the modeling and analysis of convergent systems-of-systems.  To demonstrate the methodological developments, the resilience analysis is conducted on a hypothetical energy-water nexus system of moderate size.    
\liinesfig{resilfig}{Conceptual Representation of Resilient Performance of an Engineering System\cite{Farid:2015:ISC-J19}}{Fig:resilfiig}{2.5in}

\vspace{-0.1in}
\subsection{Paper Scope \& Outline}
This work considers system-of-systems as a type of engineering system and adopts both within its scope.  
\begin{defn}[Engineering System \cite{De-Weck:2011:00}]\label{Defn:EngineeringSystem}
1.) A class of systems characterized by a high degree of technical complexity, social intricacy, and elaborate processes aimed at fulfilling important functions in society.  2.) The term engineering systems is also used to refer to the engineering discipline that designs, analyzes, verifies, and validates engineering systems.  
\end{defn}
\begin{defn}[System-of-Systems]\label{Defn:SoS}
1.) System of interest whose system elements are themselves systems; typically, these entail large-scale interdisciplinary problems with multiple, heterogeneous, distributed systems\cite{SE-Handbook-Working-Group:2015:00}. 2.) Set of systems or system elements that interact to provide a unique capability that none of the constituent systems can accomplish on its own\cite{ISO-IEC-IEEE:2019:00}.
\end{defn}

The remainder of this paper is organized as follows.  Section \ref{Sec:Graph} provides a background to graph theoretic preliminaries in terms of basic definitions, theorems, and limitations of formal graph theory and multi-layer networks.  Next, the quantification of resilience is recognized as an indirect measurement process depicted in Fig. \ref{Fig:measurement}. Consequently,  Sec. \ref{Sec:HFGT} introduces the hetero-functional graph theory model and its constituent measureables and measurement methods.  Sec. \ref{Sec:Resilience} then uses this model to develop resilience measures for convergent systems-of-systems.  These developments are then demonstrated on a hypothetical energy-water nexus system as a type of systems-of-systems.  Finally, Section \ref{Sec:Conclusion} concludes the work.  

\liinesfig{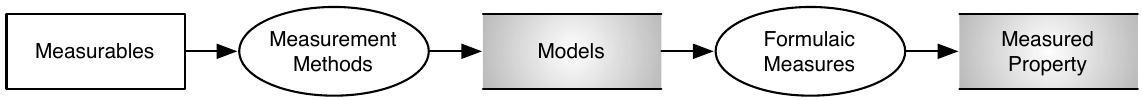}{A Generic Indirect Measurement Process\cite{Farid:2007:IEM-TP00,Farid:2017:IEM-J13}}{Fig:measurement}{3.5in}

\section{Background:  Graph Theory Preliminaries}\label{Sec:Graph}
The resilience analysis presented in this work is predicated on hetero-functional graph theory which in turn is founded on graph theory.  This section provides several definitions and theorems from graph theory in Sec. \ref{Sec:DefsAndThms}.  It then briefly explains the limitations of formal graph theory and multi-layer networks in the context of convergent systems-of-systems in Sec. \ref{Sec:FormalGraphTheory} and Sec. \ref{Sec:Multi-LayerNetworks} respectively.  

\vspace{-0.1in}
\subsection{Basic Definitions and Theorems}\label{Sec:DefsAndThms}
As mentioned in the introduction, many works in the resilience measurement literature have been based on graph theory\cite{VanBreda:2001:00,Bhamra:2011:00,Righi:2015:00,Hosseini:2016:01,Yodo:2016:02,Patriarca:2018:00,Naghshbandi:2020:00,Penaloza:2020:00,Wied:2020:00}.  A number of basic definitions and theorems from this field are introduced to support the remainder of the discussion.  
\begin{defn}
A graph\cite{Steen:2010:00}: $G=\{V,E\}$, consists of a collection of nodes $V$ and a collection of edges $E$.  Each edge $e\in E$ is said to join two nodes which are called its end points.  If $e$ joins $v_1,v_2\in V$, we write $e=\langle v_1,v_2\rangle$.  Nodes $v_1$ and $v_2$, in this case, are said to be adjacent.  Edge $e$ is said to be incident with nodes $v_1$ and $v_2$ respectively.  
\end{defn}
\begin{defn}
A directed graph (digraph)\cite{Steen:2010:00}: $G_D$, consists of a collection of nodes $V$ and a collection of arcs $A$, for which $G_D=\{V,A\}$.  Each arc $a=\langle v_1,v_2\rangle$ is said to join node $v_1\in V$ to another (not necessarily distinct) node $v_2$.  Vertex $v_1$ is called the tail of a, whereas $v_2$ is its head.  
\end{defn}
\begin{defn}\label{Defn:Bipartite}
Bipartite graph \cite{Steen:2010:00}:  A graph $G=\{V,E\}$ where $V=V_1 \cup V_2$ is bipartite if $G$ can be partitioned into two disjoint subsets $V_1$ and $V_2$ such that each edge $e \in E$ has one end point in $V_1$ and the other in $V_2$.  $E(G) \subseteq \{e = ⟨v_1,v_2⟩ | v_1 \in V1, \mbox{and } v_2 \in V_2\}$. 
\end{defn}
\begin{defn}
Path \cite{Newman:2009:00}:  Given a graph $G=\{V,E\}$, a $(v_o,v_k)$ path is an alternating sequence $[v_0, e_1, v_1, e_2, \ldots, v_{k-1}, e_k, v_k ]$ 
\end{defn}
\begin{defn}
Incidence matrix\cite{Steen:2010:00}: $M=M^+-M^-$ of size $|V|\times|A|$ is given by:
\begin{align}
M^+(i,j)&=\left\{ 
\begin{array}{ll}
 1 & \mbox{if vertex $v_i$ is the tail of arc $a_j$} \\
 0 & \mbox{otherwise}
\end{array}
\right. \\
M^-(i,j)&=\left\{ 
\begin{array}{ll}
 1 & \mbox{if vertex $v_i$ is the head of arc $a_j$} \\
 0 & \mbox{otherwise}
\end{array}
\right.
\end{align}
\end{defn} 
\begin{defn}
Adjacency matrix\cite{Steen:2010:00}: $A$, is binary and of size $|V|\times|V|$ and its elements are given by
\begin{align}
A(i,j)&=\left\{ 
\begin{array}{ll}
1 & \mbox{if $\langle v_i,v_j\rangle$ exists} \\
0 & \mbox{otherwise}
\end{array}
\right. \\
A&=M^{+}M^{-T}
\end{align}
\end{defn}
\begin{thm}\label{Th:NumPaths}
Number of Paths in a Graph\cite{Newman:2009:00}: The number of k-node (or k-1 edge-step) paths $|\mathds{P}_k|$ between nodes i and j in a graph is given by $A^{(k-1)}(i,j)$.
\end{thm}
\begin{thm}\label{Th:NumLoops}
Number of Loops in a Graph\cite{Newman:2009:00}:  The number of k-node (or k-1 edge-step) paths loops from node i back to itself is given by $A^{(k-1)}(i,i)$.  
\end{thm}

\vspace{-0.1in}
\subsection{Limitations of Formal Graph Theory}\label{Sec:FormalGraphTheory}
While for decades, graph theory has presented a useful abstraction across many applications, it has limitations in the systems engineering of convergent systems-of-systems.  Graph theory, as it is traditionally applied, focuses primarily on an abstracted model of a system's form; neglecting an explicit description of system's function\cite{De-Weck:2011:00,Crawley:2015:00}.  Newman lists common applications of (formal) graph theory in Table \ref{Ta:Networks}.  In all cases, nodes and edges in a formal graph represents nouns; with nodes typically representing point objects in space and edges representing line objects.  The system function, what a system does, in terms of verbal phrases, has been entirely omitted from the \emph{explicit} statement of the formal graph and any understanding of the system's function is implicit.  Consequently, all of the applications listed in Table \ref{Ta:Networks} describe \emph{homo-functional} engineering systems where operands (of some specific type) are transported between physical locations.  While transportation processes that account for movement from one location to another are fundamentally different, ultimately, they are of the same class or type. Thus, it is less than clear how formal graph theory may be applied to convergent systems-of-systems that are of a fundamentally transformative nature with multiple operands.  As convergent systems-of-systems are \emph{hetero-functional} (i.e. include functions and operands of many types), formal graph theory may impede rigorous approaches where resilience can be engineered into the system.  
\vspace{-0.15in}
\begin{table}[htbp]
\begin{center}
\caption{Common Applications of Graph Theory\cite{Newman:2009:02624}}\label{Ta:Networks}
\vspace{-4pt}
%\begin{scriptsize}
\begin{tabular}{lll}\hline
System/Network & Node & Edge \\
Internet & Computer/Router & Wired/Wireless Data Connection\\
World Wide Web & Web Page & Hyperlink \\
Power Grid & Power or Substation & Transmission Line \\
Transportation & Intersections & Roads \\
Neural Network & Neuron & Synapse \\\hline
\end{tabular}
%\end{scriptsize}
\end{center}
\vspace{-0.2in}
\end{table}

\vspace{-0.1in}
\subsection{Limitations of Multi-Layer Networks}\label{Sec:Multi-LayerNetworks}
To address the inherent complexity in many engineering systems, the field of network science has generalized formal graphs into multi-layer networks\cite{DAgostino:2014:00,Kivela:2014:00}.  In a recent comprehensive review Kivela et. al \cite{Kivela:2014:00} write:

\begin{quoting}
``The study of multi-layer networks $\ldots$ has become extremely popular.  Most real and engineered systems include multiple subsystems and layers of connectivity and developing a deep understanding of multi-layer systems necessitates generalizing `traditional' graph theory.  Ignoring such information can yield misleading results, so new tools need to be developed.  One can have a lot of fun studying `bigger and better' versions of the diagnostics, models and dynamical processes that we know and presumably love -- and it is very important to do so but the new `degrees of freedom' in multi-layer systems also yield new phenomena that cannot occur in single-layer systems.  Moreover, the increasing availability of empirical data for fundamentally multi-layer systems amidst the current data deluge also makes it possible to develop and validate increasingly general frameworks for the study of networks.  

$\ldots$ Numerous similar ideas have been developed in parallel, and the literature on multi-layer networks has rapidly become extremely messy.  Despite a wealth of antecedent ideas in subjects like sociology and engineering, many aspects of the theory of multi-layer networks remain immature, and the rapid onslaught of papers on various types of multilayer networks necessitates an attempt to unify the various disparate threads and to discern their similarities and differences in as precise a manner as possible.

$\ldots$ [The multi-layer network community] has produced an equally immense explosion of disparate terminology, and the lack of consensus (or even generally accepted) set of terminology and mathematical framework for studying is extremely problematic."
\end{quoting}
In addition to the above limitations, Kivela et. al showed that \emph{all} of the reviewed works have exhibited at least one of eight different types of modeling constraints\cite{Kivela:2014:00}.  To demonstrate the consequences of these modeling limitations, the HFGT text\cite{Schoonenberg:2019:ISC-BK04} developed a small, but highly heterogeneous, hypothetical test case system that exhibited \emph{all eight} of the modeling limitations identified by Kivela et. al.  Consequently, none of the multi-layer network models identified by Kivela et. al. would be able to model such a hypothetical test case.   In contrast, a complete HFGT analysis of this hypothetical test case was demonstrated in the aforementioned text\cite{Schoonenberg:2019:ISC-BK04}.  To follow up this result, the tensor formulation of hetero-functional graph theory proved that multi-layer networks are neither ontologically lucid nor complete\cite{Farid:2022:ISC-J51}.  

\vspace{-0.1in}
\section{Hetero-functional Graph Theory Model}\label{Sec:HFGT}
Hetero-functional graph theory overcomes the limitations in formal graph theory and multi-layer networks through its connection to model-based systems engineering \cite{Crawley:2015:00} which in turn is founded in the universal structure of human language \cite{Cook:2014:01}.  Both of these connections help enhance \emph{convergence potential} of HFGT as it addresses systems-of-systems.  More specifically, HFGT includes an explicit description of a system's form, function, and the allocation of the latter on to the former.  This dichotomy of form and function is repeatedly emphasized in the fields of engineering design and systems engineering\cite{Crawley:2015:00,Buede:2009:00,Kossiakoff:2003:00,Farid:2016:ISC-BK03}.  Consequently, HFGT has been able to model production\cite{Farid:2008:IEM-J04,Farid:2008:IEM-J05,Farid:2008:IEM-J06,Farid:2015:IEM-J23,Farid:2017:IEM-J13}, transportation\cite{Viswanath:2013:ETS-J08}, electric power\cite{Farid:2015:SPG-J17,Thompson:2021:SPG-J46}, healthcare\cite{Farid:2015:SPG-J17,Khayal:2017:ISC-J35,Khayal:2021:ISC-J48}, multi-modal electrified transportation\cite{Farid:2016:ETS-J27,vanderWardt:2017:ETS-J33}, microgrid-enabled production\cite{Schoonenberg:2017:IEM-J34}, integrated hydrogen natural gas\cite{Schoonenberg:2022:ISC-J50}, multi-energy\cite{Thompson:2024:ISC-J55}, and interdependent smart city infrastructure systems\cite{Schoonenberg:2019:ISC-BK04}.  This section introduces hetero-functional graph theory to support the resilience analysis methodology in the following section.    

\vspace{-0.1in}
\subsection{System Resources, Processes, Operands}\label{sec:JS}
Hetero-functional graph theory makes its connection to human language explicit through a set of system resources $R$ as subjects, a set of system processes $P$ as predicates, and a set of operands $L$ as their constituent objects.  

\begin{defn}[System Resource\cite{SE-Handbook-Working-Group:2015:00}]\label{Defn:Resource}
An asset or object $r_v \in R$ that is utilized during the execution of a process.  
\end{defn}

\begin{defn}[System Process\cite{Hoyle:1998:00,SE-Handbook-Working-Group:2015:00}]\label{Defn:Process}
An activity $p_w \in P$ that transforms a predefined set of input operands into a predefined set of outputs. 
\end{defn}

\begin{defn}[System Operand\cite{SE-Handbook-Working-Group:2015:00}]\label{Defn:Operand}
An asset or object $l_i \in {\cal L}$ that is operated on or consumed during the execution of a process.  
\end{defn}

Hetero-functional graph theory further recognizes that there are inherent differences within the set of resources as well as within the set of processes.   Therefore, as shown in Fig. \ref{Fig:HFGT-Meta}, classifications of these sets of resources and sets of processes are introduced.  $R=M \cup B \cup H$ where $M$ is the set of transformation resources, $B$ is the set of independent buffers, and $H$ is the set of transportation resources.   Furthermore, the set of buffers $B_S=M \cup B$ is introduced for later discussion.  Similarly, $P = P_\mu \cup P_{\bar{\eta}}$ where $P_\mu$ is the set of transformation processes  and $P_{\bar{\eta}}$ is the set of refined transportation processes.  The latter, in turn, is determined from the Cartesian product (\Cross) of the set of transportation processes $P_\eta$ and the set of holding processes $P_\gamma$.  $P_{\bar{\eta}} = P_{\gamma} \mbox{\Cross} P_{\eta}$.  Further explanation of HFGT meta-architecture and its taxonomies of processes and resources can be found in \cite{Schoonenberg:2019:ISC-BK04}.  

\liinesbigfig{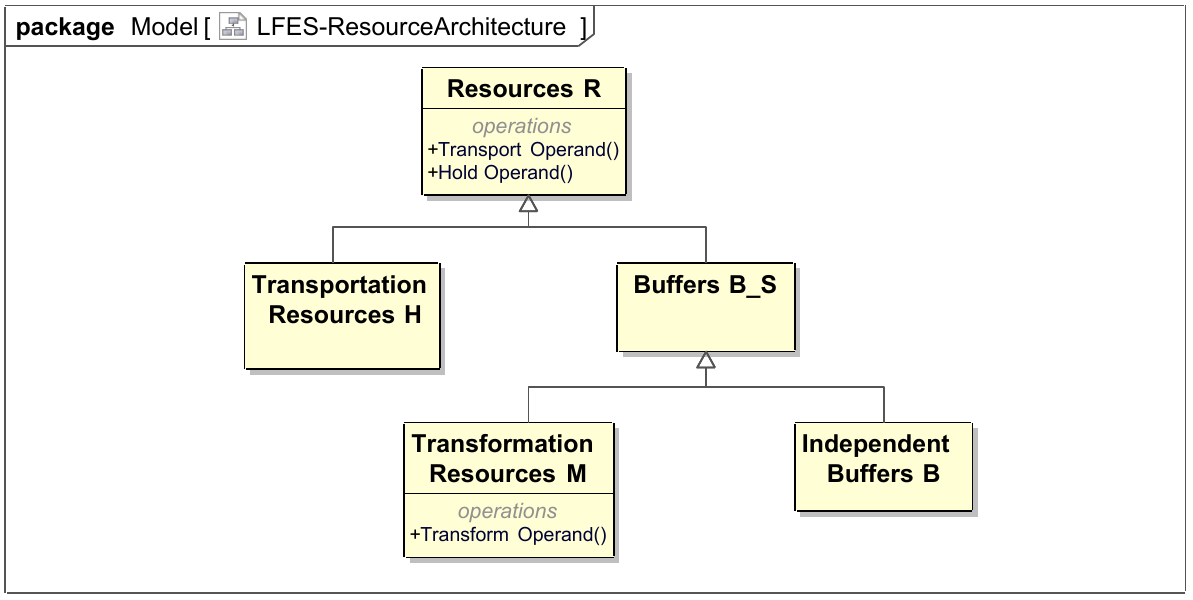}{The Hetero-functional Graph Theory Meta-Architecture drawn using the Systems Markup Language (SysML).  It consists of three types of resources $R = M \cup B \cup H$ that are capable of two types of process $P_{\bar{\eta}} = P_{\gamma} \mbox{} P_{\eta}$\cite{Schoonenberg:2019:ISC-BK04}.}{HFGT-Meta}

\vspace{-0.1in}
\subsection{Existence, Availability, and Concept of System Capabilities}\label{sec:AS}
System processes can be allocated to system resources to form subject+verb+object sentences called system capabilities.    
\begin{defn}[Capability\cite{Schoonenberg:2019:ISC-BK04,Farid:2022:ISC-J51,Farid:2016:ISC-BC06}]\label{Defn:Capability}
An action $\epsilon_{wv} \in {\cal E}_S$ (in the SysML sense) defined by a system process $p_w \in P$ being executed by a resource $r_v \in R$.  It constitutes a subject + verb + operand sentence of the form: ``Resource $r_v$ does process $p_w$".  
\end{defn} 
\noindent From a systems engineering perspective, and on a system level, the allocation of system processes to system resources is captured in the ``design equation"\cite{Schoonenberg:2019:ISC-BK04,Farid:2022:ISC-J51,Farid:2016:ISC-BC06}: 
\begin{equation}\label{Eq:designequation}
P=A_S\odot R
\end{equation}
where $A_S$ is the system knowledge base, and $\odot$ is matrix Boolean multiplication.  
\begin{defn}[System Concept\cite{Schoonenberg:2019:ISC-BK04,Farid:2022:ISC-J51,Farid:2016:ISC-BC06}]\label{Defn:SystemConcept}
A binary matrix $A_S$ of size $|P| \times |R|$ whose element $A_S(w,v)\in\{0,1\}$ is equal to one when action $e_{wv} \in {\cal E}_S$ (in the SysML sense) is available as a system process $p_w \in P$ being executed by a resource $r_v \in R$.
\begin{equation}
A_S=J_S\ominus K_S=J_S\cdot \neg{K}_S
\end{equation}
where $\ominus$ is Boolean subtraction, $\cdot$ is the Hadamard product, and $\neg{K}_S=NOT(K_S)$.  
\end{defn} 
\noindent In other words, the system concept forms a bipartite graph (Defn. \ref{Defn:Bipartite}) between the set of system processes and the set of system resources.  Additionally, hetero-functional graph theory differentiates between the \emph{existence} and the \emph{availability} of physical capabilities in the system.  While the former is described by the system knowledge base $J_S$, the latter is captured by the system constraints matrix $K_S$(which is assumed to evolve in time)\cite{Schoonenberg:2019:ISC-BK04,Farid:2022:ISC-J51,Farid:2016:ISC-BC06}.  
\begin{defn}[System Knowledge Base\cite{Schoonenberg:2019:ISC-BK04,Farid:2022:ISC-J51,Farid:2016:ISC-BC06}]\label{Defn:KnowledgeBase}
A binary matrix $J_S$ of size $|P| \times |R|$ whose element $J_S(w,v)\in\{0,1\}$ is equal to one when action $e_{wv} \in {\cal E}_S$ (in the SysML sense) exists as a system process $p_w \in P$ being executed by a resource $r_v \in R$.     
\end{defn} 
\begin{defn}[System Constraints Matrix\cite{Schoonenberg:2019:ISC-BK04,Farid:2022:ISC-J51,Farid:2016:ISC-BC06}]\label{Defn:SystemConstraint}  A binary matrix $K_S$ of size $|P| \times |R|$ whose element $K_S(w,v)\in\{0,1\}$ is equal to one when a constraint eliminates event $e_{wv}$ from the event set.  
\end{defn}

Every filled element of the system concept indicates a \emph{system capability} (Defn. \ref{defn:capability}) of the form:  ``Resource $r_v$ does process $p_w$".   The system constraints matrix limits the availability of capabilities in the system knowledge base to create the system concept  $A_S$.  Importantly, in the next section, the set of capabilities ${\cal E}_S$ become the nodes of a hetero-functional graph adjacency matrix.  

As the system concept $A_S$ is very sparse, it is often computationally useful to to introduce a (non-unique) projection operator $\mathds{P}_S$ that eliminates this sparsity by projecting the system knowledge base onto a one's vector\cite{Schoonenberg:2019:ISC-BK04,Farid:2022:ISC-J51,Farid:2016:ISC-BC06}.  
\begin{align}
\widetilde{J}_S&=\mathds{P}_S J_S^V = \mathds{1}^{|{\cal E}_S|}\\ \label{eq:projHFAM}
\widetilde{K}_S&=\mathds{P}_S K_S^V \\
\widetilde{A}_S&=\mathds{P}_S A_S^V
\end{align}
where $()^V$ is shorthand for matrix vectorization (i.e. $vec()$).  In such a case, a system capability $\epsilon_{wv} \in {\cal E}_S$ becomes more straightforwardly labelled by the index  $\psi \in \left[1, \dots ,|{\cal E}_S|\right]$.     

\vspace{-0.1in}
\subsection{Hetero-functional Incidence Tensor and Engineering System Net}\label{sec:Mrho}
Once the system capabilities ${\cal E}_S$ have been defined within the system concept $A_S$, they can be related to one another through the hetero-functional incidence tensor and subsequently simulated with the engineering system net.  The (third-order) hetero-functional hetero-functional incidence tensor $\widetilde{\cal M}_\rho$ describes the structural relationships between the physical capabilities ${\cal E}_S$, the system operands $L$, and the system buffers $B_S$.   

\begin{equation}
\widetilde{\cal M}_\rho=\widetilde{\cal M}_\rho^+-\widetilde{\cal M}_\rho^-
\end{equation}
\begin{defn}[The Negative 3$^{rd}$ Order Hetero-functional Incidence Tensor $\widetilde{\cal M}_\rho^-$ \cite{Farid:2022:ISC-J51}]\label{Defn:MRhoNeg}
The negative hetero-functional incidence tensor $\widetilde{\cal M_\rho}^- \in \{0,1\}^{|{\cal L}|\times |B_S| \times |{\cal E}_S|}$  is a third-order tensor whose element $\widetilde{\cal M}_\rho^{-}(i,y,\psi)=1$ when the system capability ${\epsilon}_\psi \in {\cal E}_S$ pulls operand $l_i \in {\cal L}$ from buffer $b_{s_y} \in B_S$.
\end{defn} 
\begin{defn}[The Positive  3$^{rd}$ Order Hetero-functional Incidence Tensor $\widetilde{\cal M}_\rho^+$\cite{Farid:2022:ISC-J51}]\label{Defn:MRhoPos}
The positive hetero-functional incidence tensor $\widetilde{\cal M}_\rho^+ \in \{0,1\}^{|L|\times |B_S| \times |{\cal E}_S|}$  is a third-order tensor whose element $\widetilde{\cal M}_\rho^{+}(i,y,\psi)=1$ when the system capability ${\epsilon}_\psi \in {\cal E}_S$ injects operand $l_i \in {\cal L}$ into buffer $b_{s_y} \in B_S$.
\end{defn} 

In order to facilitate matrix-based calculation, the third-order hetero-functional incidence tensor $\widetilde{\cal M}_\rho$ can be matricized into its associated second order form $\widetilde{M}_\rho$ (where the first two dimensions are combined into a single dimension).  The resulting matrix has a size of $|{\cal L}||B_S| \times |\cal{E_S}|$\cite{Farid:2022:ISC-J51}.  The underlying physical intuition provides each buffer $b_{s_{y}} \in B_S$ with $|{\cal L}|$ places (i.e. bins or shelves) for each operand at that buffer.  These $|{\cal L}||B_S|$ places form a bipartite graph with the system's physical capabilities ${\cal E_S}$.  

Consequently, the supply, demand, transportation, storage, transformation, assembly, and disassembly of multiple operands in distinct locations over time can be described by an Engineering System Net and its associated State Transition Function\cite{Schoonenberg:2022:ISC-J50}.  
\begin{defn}[Engineering System Net\cite{Schoonenberg:2022:ISC-J50}]\label{Defn:ESN}
An elementary Petri net ${\cal N}_S = \{S, {\cal E}_S, \textbf{M}, W, Q\}$, where
\begin{itemize}
\item $S$ is the set of engineering system places with size: $|{\cal L}||B_S|$,
\item ${\cal E}_S$ is the set of engineering system transitions with size: $|{\cal E}_S|$,
\item $\textbf{M} \subseteq (S \times {\cal E}_{S}) \cup ({\cal E}_{S} \times S)$ is the set of arcs, with the associated incidence matrices: $\widetilde{M}_\rho = \widetilde{M}^+_\rho - \widetilde{M}^-_\rho$,
\item $W$ a the set of weights on the arcs may be added to the nonzero elements of the incidence matrices $\widetilde{M}_\rho^+,\widetilde{M}_\rho^-$,
\item $Q=[Q_B; Q_E]$ is the marking vector for both the set of places and the set of transitions. 
\end{itemize}
\end{defn}
\noindent Without loss of generality, the engineering system net is assumed to begin at a single transition (with no preset places) so as to represent the initiation of engineering system net operation.  Similarly, the engineering system net is assumed to end at a single transition (with no postset places) so as to represent the termination of engineering system operation.  
\begin{defn}[Engineering System Net State Transition Function\cite{Schoonenberg:2022:ISC-J50}]\label{Defn:ESN-STF}
The  state transition function of the engineering system net $\Phi_S()$ is:
\begin{equation}\label{CH6:eq:PhiCPN}
Q[k+1]=\Phi_S\left(Q[k],U_S^-[k], U_S^+[k]\right) \quad \forall k \in \{1, \dots, K\}
\end{equation}
where $k$ is the discrete time index, $K$ is the simulation horizon, $Q=[Q_{B}; Q_{\cal E}]$, $Q_B$ has size $|{\cal L}||B_S| \times 1$, $Q_{\cal E}$ has size $|{\cal E}_S|\times 1$, the input firing vector $U_S^-[k]$ has size $|{\cal E}_S|\times 1$, and the output firing vector $U_S^+[k]$ has size $|{\cal E}_S|\times 1$.  
\begin{align}\label{CH6:CH6:eq:Q_B:HFNMCFprogram}
Q_{B}[k+1]&=Q_{B}[k]+\widetilde{M}_\rho^+U_S^+[k]\Delta T-\widetilde{M}_\rho^-U_S^-[k]\Delta T \\ \label{CH6:CH6:eq:Q_E:HFNMCFprogram}
Q_{\cal E}[k+1]&=Q_{\cal E}[k]-U^+[k]\Delta T +U_S^-[k]\Delta T
\end{align}
where $\Delta T$ is the duration of the simulation time step.  
\end{defn}
\noindent Furthermore, the presence of nonzero elements in the system constraints matrix $K_S$ forces the associated elements of the engineering system net firing vectors to zero.  
\begin{align}\label{Eq:USConstraint}
\widetilde{K}_S^TU_S^-[k] &=0 \\\nonumber
\widetilde{K}_S^TU_S^+[k] &=0
\end{align}
For the sake of simplicity in this work, and without loss of generality, the transitions ${\cal E}_{S}$ are assumed to occur instantaneously.  $U_{S}^+[k]=U_{S}^-[k] = U_{S}[k] \, \forall k \in \{1, \dots, K\}$, at which point Eq. \ref{CH6:CH6:eq:Q_E:HFNMCFprogram} collapses to triviality.  

\vspace{-0.1in}
\subsection{Hetero-functional Adjacency Matrix}
The construction of the second-order hetero-functional incidence matrix $\widetilde{M}_{\rho}=\widetilde{M}_{\rho}^+ + \widetilde{M}_{\rho}^-$ facilitates the creation of a hetero-functional adjacency matrix $A_\rho$ to represent the feasible pairwise sequences of capabilities\cite{Schoonenberg:2019:ISC-BK04,Farid:2022:ISC-J51,Farid:2016:ISC-BC06}.  
\begin{defn}[Hetero-functional Adjacency Matrix\cite{Schoonenberg:2019:ISC-BK04,Farid:2022:ISC-J51,Farid:2016:ISC-BC06}]\label{Defn:ARho}
A square binary matrix $A_\rho$ of size $|R|P| \times |R||P|$ whose element $A_\rho(\chi_1,\chi_2)\in \{0,1\}$ is equal to one when string $z_{\chi_1,\chi_2}=\epsilon_{w_1v_1}\epsilon_{w_2v_2} \in {\cal Z}$ is available and exists, where index $\chi_i \in \left[1, \dots , |R||P|\right]$.
\begin{align}
A_\rho = J_\rho \ominus K_\rho
\end{align}
\end{defn}
\noindent In other words, the hetero-functional adjacency matrix corresponds to a hetero-functional graph $G = \{{\cal E}_S, {\cal Z} \}$ with capabilities ${\cal E}_S$ as nodes and feasible sequences ${\cal Z}$ as edges.  Again, it is often computationally useful to eliminate the sparsity of zero rows and columns in $A_\rho$ and instead use the projected hetero-functional adjacency matrix $\widetilde{A}_{\rho}$\cite{Farid:2022:ISC-J51}.  
\begin{align}
\widetilde{A}_{\rho} = \mathds{P}_S A_\rho \mathds{P}_S^T
\end{align}
Consequently, the projected hetero-functional adjacency matrix $\widetilde{A}_{\rho}$ can be calculated as a matrix product of the positive and negative hetero-functional incidence matrices $\widetilde{M}_\rho^+$ and $\widetilde{M}_\rho^+$\cite{Farid:2022:ISC-J51}.  
 \begin{equation}\label{eq:bigresult}
 \widetilde{A}_{\rho} = \widetilde{M}_\rho^{+T}\odot \widetilde{M}_\rho^- = \widetilde{M}_\rho^{+T}\widetilde{M}_\rho^-
 \end{equation}
where the Boolean and real matrix products are interchangeable because each process is associated with exactly one origin-destination pair\cite{Farid:2022:ISC-J51}.    

\vspace{-0.1in}
\subsection{Operand Behavior}\label{Sec:OperandBehavior}
In addition to identifying the potential for multiple operands (Defn. \ref{Defn:Operand}) in a system-of-systems, hetero-functional graph theory recognizes that each operand may have a behavior whose state that evolves in time.  In the case of commodities, the operand may simply come in and out of existence (e.g. by virtue of a chemical process).  In more complex products, raw material may be transformed into work-in-progress, and then again into final products.  Operands may also represent services at various stages of delivery.  Finally, operands may also represent complex services that consist of multiple operands, products and services and their associated state evolution.  The state evolution of each operand $l_i \in {\cal L}$ is described by an Operand Net ${\cal N}_{l_i}$ and its associated state transition function $\Phi_{l_i}()$.  
\begin{defn}[Operand Net\cite{Farid:2008:IEM-J04,Schoonenberg:2019:ISC-BK04,Khayal:2017:ISC-J35,Schoonenberg:2017:IEM-J34}]\label{Defn:OperandNet} 
Given operand $l_i \in {\cal L}$, an elementary Petri net ${\cal N}_{l_i}= \{S_{l_i}, {\cal E}_{l_i}, \textbf{M}_{l_i}, W_{l_i}, Q_{l_i}\}$ where 
\begin{itemize}
\item $S_{l_i}$ is the set of places describing the operand's state.  
\item ${\cal E}_{l_i}$ is the set of transitions describing the evolution of the operand's state.
\item $\textbf{M}_{l_i} \subseteq (S_{l_i} \times {\cal E}_{l_i}) \cup ({\cal E}_{l_i} \times S_{l_i})$ is the set of arcs, with the associated incidence matrices: $M_{l_i} = M^+_{l_i} - M^-_{l_i}$.  
\item $W_{l_i} : \textbf{M}_{l_i}$ is the set of weights on the arcs, as captured in the incidence matrices $M^+_{l_i},M^-_{l_i}$.  
\item $Q_{l_i}= [Q_{Sl_i}; Q_{{\cal E}l_i}]$ is the marking vector for both the set of places and the set of transitions. 
\end{itemize}
\end{defn}
\noindent Without loss of generality, the operand net is assumed to begin at a single transition (with no preset places) so as to represent an operand coming into existence.  Similarly, the operand net is assumed to end at a single transition (with no postset places) so as to represent an operand disappearing from existence.  

\begin{defn}[Operand Net State Transition Function\cite{Schoonenberg:2019:ISC-BK04,Khayal:2017:ISC-J35,Schoonenberg:2017:IEM-J34}]\label{Defn:OperandNet-STF}
The  state transition function of each operand net $\Phi_{l_i}()$ is:
\begin{equation}\label{CH6:eq:PhiSPN}
Q_{l_i}[k+1]=\Phi_{l_i}\left(Q_{l_i}[k],U_{l_i}^-[k], U_{l_i}^+[k]\right) \quad \forall k \in \{1, \dots, K\} 
\end{equation}
where $Q_{l_i}=[Q_{Sl_i}; Q_{{\cal E} l_i}]$, $Q_{Sl_i}$ has size $|S_{l_i}| \times 1$, $Q_{{\cal E} l_i}$ has size $|{\cal E}_{l_i}| \times 1$, the input firing vector $U_{l_i}^-[k]$ has size $|{\cal E}_{l_i}|\times 1$, and the output firing vector $U^+[k]$ has size $|{\cal E}_{l_i}|\times 1$ and $K$ is the event horizon of the operand (net).  

\begin{align}\label{CH6:CH6:eq:Q_B:HFNMCFprogram}
Q_{Sl_i}[k+1]&=Q_{Sl_i}[k]+{M^+_{l_i}}U_{l_i}^+[k]\Delta T - {M^-_{l_i}}U_{l_i}^-[k]\Delta T \\ \label{CH6:CH eq:Q_E:HFNMCFprogram}
Q_{{\cal E} l_i}[k+1]&=Q_{{\cal E} l_i}[k]-U_{l_i}^+[k]\Delta T +U_{l_i}^-[k]\Delta T
\end{align}
\end{defn}
\noindent For the sake of simplicity in this work, and without loss of generality, the transitions ${\cal E}_{l_i}$ are assumed to occur instantaneously.  $U_{l_i}^+[k]=U_{l_i}^-[k] = U_{l_i}[k] \, \forall k \in \{1, \dots, K\}$, at which point Eq. \ref{CH6:CH eq:Q_E:HFNMCFprogram} collapses to triviality.  

\vspace{-0.1in}
\subsection{Operand Net Feasibility}
Finally, hetero-functional graph theory recognizes that any given transition in an operand net cannot be realized without the execution of a corresponding transition in the engineering system net.  This correspondence is described by the operand-capability feasibility matrix.  
\begin{defn}[Operand-Capability Feasibility Matrix\cite{Schoonenberg:2022:ISC-J50}]\label{Defn:SyncMat}
For a given operand $l_i$, a binary matrix of size $|{\cal E}_{l_i}| \times |{\cal E}_S|$ whose value $\widetilde{\Lambda}_{i}(x,\psi)=1$ if $\epsilon_{xl_i}$ can be feasibly realized by capability $\epsilon_{s\psi}$. 
\end{defn}
\noindent In the context of this paper, without loss of generality, the operand-capability feasibility matrix $\Lambda_i$ is assumed to have a single nonzero element in the row associated with the first transition of the operand net.  Similarly, it is assumed to have a single nonzero element in the row associated with the last transition of the operand net.  The operand-capability feasibility matrix allows the engineering system net firing vectors to evolve the operand net firing vectors\cite{Schoonenberg:2022:ISC-J50}.  
\begin{align}\label{Eq:Operand-Capability-Feasibility-Condition2}
U_{l_i}[k] & = \widetilde{\Lambda}_i U_S[k]    \quad \forall k \in \{1, \dots, K\}
\end{align}
Consequently, the presence of a nonzero elements in the engineering system net firing vector requires that there exist at least one nonzero element in the operand net firing vector\cite{Schoonenberg:2019:ISC-BK04}.  
\begin{align}\label{Eq:Operand-Capability-Feasibility-Condition}
bi\big(U_S[k]\big)& = \bigvee_i^{|{\cal L}|} \widetilde{\Lambda}_i^T bi\big(U_{l_i}[k]\big)    \quad \forall k \in \{1, \dots, K\}
\end{align}
where $bi()$ is the function that returns one for all nonzero values and zero otherwise.  

\vspace{-0.1in}
\section{Development of Resilience Measures}\label{Sec:Resilience}
The introduction to hetero-functional graph theory in the previous section provides a strong foundation for a resilience analysis methodology in this section.  Many resilience measures in the existing literature are based upon some form of calculation of the shortest path length through a formal graph\cite{Holme:2002:00,Ash:2007:00,Ip:2011:00,Albert:2000:00}.  In contrast, the resilience measures developed here are based upon the \emph{number} rather than the \emph{length} of graph paths.  Furthermore, the graphs presented here rely on a hetero-functional rather than a formal graph.  Consequently, Sec. \ref{Sec:PathEnum} enumerates the number of paths for an operand flowing through a hetero-functional adjacency matrix of an engineering system.  Sec. \ref{Sec:Performance} then defines the performance of an engineering system.   Sections \ref{Sec:AER}, \ref{Sec:LER}, and \ref{Sec:DER} then define latent, actual, and dynamic engineering resilience measures respectively.  The resilience measures developed in this section a formal generalization of previous work\cite{Farid:2015:ISC-J19,Farid:2014:ISC-C37,Farid:2014:ISC-C38,Thompson:2020:SPG-C68,Thompson:2021:SPG-J46} because they address operand nets of arbitrary topology and not just those with a single sequence of places and transitions.  

\vspace{-0.1in}
\subsection{Path Enumeration in Engineering Systems}\label{Sec:PathEnum}
From Theorem \ref{Th:NumPaths}, the number of k-capability paths through a hetero-functional graph between an arbitrary pair of capabilities is given by $\widetilde{A}_{Pk}=\widetilde{A}_\rho^{(k-1)}$ where $\widetilde{A}_\rho$ is the projected hetero-functional adjacency matrix (Defn. \ref{Defn:ARho}).  As self-loops are assumed not to contribute to resilience, the number of \emph{simple} k-capability paths (i.e. without loops) between an arbitrary pair of capabilities $\widetilde{A}_{Pk}$ is calculated recursively from Theorem \ref{Th:NumLoops}.  
\begin{align}
\widetilde{A}_{Pk} &=[\widetilde{A}_{Pk-1}][\widetilde{A}_{\rho s}]\\
\widetilde{A}_{\rho s}&= \widetilde{A}_{\rho} - diag(\widetilde{A}_\rho) \\
\widetilde{A}_{P2} &= \widetilde{A}_{\rho s}
\end{align}
In the context of resilience measurement, the calculation of $\widetilde{A}_{Pk}$ must be generalized so as to include the possibility that some capabilities may not be available at a given time.  Consequently, 
\begin{align}
\widetilde{A}_{Pk} &=\big[\widetilde{A}_{Pk-1}\big]*\big[\widetilde{A}_{\rho s}\cdot\big(\mathbf{1}*\neg{\widetilde{K}}_S^{T}\big)\big]\\
\widetilde{A}_{P2} &= \big[\neg{\widetilde{K}}_S*\mathbf{1}^T\big]\cdot\widetilde{A}_{\rho s}\cdot\big[\mathbf{1}*\neg{\widetilde{K}}_S^{T}\big]
\end{align}
where the projected system constraint matrix $\widetilde{K_S}$ (Defn. \ref{Defn:SystemConstraint}) introduces a binary availability condition to the formulation, $\neg$ is Boolean NOT, and $()^V$ is matrix vectorization (i.e. $vec()$).    

Next, it is important to recognize that a given operand $l_i \in {\cal L}$ (Defn. \ref{Defn:Operand}) may not require all available capabilities in a hetero-functional graph.  As an operand net $N_{l_i}$ (Defn. \ref{Defn:OperandNet}) is executed, it introduces a sequence of operand net firing vectors $U_{l_i}[k]$.  These, in turn, by Eq. \ref{Eq:Operand-Capability-Feasibility-Condition}, must find a feasible correspondence with the engineering system net firing vectors $U_{S}[k]$.  Therefore, the engineering system net firing vectors in binary form $bi(U_{S}[k])$ can be used to select out relevant parts of the hetero-functional graph.  Consequently, the calculation of $\widetilde{A}_{Pk}$ must be further generalized to account for the applicability to a given operand $l_i$.  
\begin{align}
\widetilde{A}_{Pk} &=[\widetilde{A}_{Pk-1}]*\Big[\widetilde{A}_{\rho s}\cdot\Big(\mathbf{1}*\big[bi\big(U_{S}[k]\big)\cdot\neg{\widetilde{K}}_S\big]^T\Big)\Big]\\ \label{Eq:AP2}
\widetilde{A}_{P2} &= \Big[\Big(bi\big(U_{S}[1]\big)\cdot\neg{\widetilde{K}}_S\Big)*\mathbf{1}^T\Big]\cdot\widetilde{A}_{\rho s}\cdot\Big[\mathbf{1}*\Big(bi\big(U_{S}[2]\big)\cdot\neg{\widetilde{K}}_S\Big)^T\Big]
\end{align}

This calculation of $\widetilde{A}_{Pk}$ requires the determination of the firing operand net vectors $U_{l_i}[k]$ such that it completes its behavior in a pre-specified number of steps $K$.  In many cases, operand nets either have a small size or simple topology that lend themselves to determining these firing vectors by inspection or by simple forward or backward scheduling heuristics.  However, in the general case of a complex operand net topology, a mixed integer program must be solved.  
\begin{alignat}{3}\label{Eq:ObjFunc}
\text{maximize } {\cal J}_{l_i} &= \sum_{k=1}^{K} \mathds{1}^TU_{l_i}[k] \\ \label{Eq:STF1}
\text{s.t. } Q_{Fl_i}[k]&=Q_{Sl_i}[k] - {M^-_{l_i}}U_{l_i}[k]\Delta T && \quad \forall k \in \{1, \dots, K\} \\ \label{Eq:STF2}
Q_{Sl_i}[k+1]&=Q_{Fl_i}[k] + {M^+_{l_i}}U_{l_i}[k]\Delta T && \quad \forall k \in \{1, \dots, K-1\}\\ \label{Eq:InitCond}
Q_{S_{l_i}}[1]&=K\cdot{M^+_{l_i}}e_{x1} \\ \label{Eq:InitialTransition}
e_{x1}^TU_{l_i}[k]&= 0  && \quad \forall k \in \{2, \dots, K\} \\ \label{Eq:FeasibilityMat}
U_{l_i}[k] & \leq \widetilde{\Lambda}_i \big(\neg{\widetilde{K}}_S\big)  && \quad \forall k \in \{1, \dots, K\} \\
Q_{Sl_i}[k] \in \mathbb{R}^{+|S_{l_i}|}, & Q_{Fl_i}[k] \in \mathbb{R}^{+|S_{l_i}|}, U_{l_i}[k] \in \{0,1\}^{|{\cal E}_S|}  && \quad \forall k \in \{1, \dots, K\}
\end{alignat}
where the objective function in Eq. \ref{Eq:ObjFunc} maximizes the number of operand net transitions fired over time, Eq. \ref{Eq:STF1} is the first half of the operand net state transition function that imposes a non-negativity firing condition, and Eq. \ref{Eq:STF2} is the second half of the operand net state transition function.  The initial condition on $Q_{S_{l_i}}$ in Eq. \ref{Eq:InitCond} ensures that there is a large number of tokens in the postset places of the initial transition $\epsilon_{x1}$.  These tokens serve to enable the evolution of the operand net through Eq. \ref{Eq:STF1}.  (Note that, $e_{x1}$ is the $x_1^{th}$ elementary basis vector of appropriate size.). Furthermore, Eq. \ref{Eq:InitialTransition} ensures that the initial transition is not fired after the first time step.    Finally, because the system constraints matrix limits the nonzero elements of the engineering system net firing vectors (via Eq. \ref{Eq:USConstraint}), and they, in turn, evolve the operand net firing vectors (via Eq. \ref{Eq:Operand-Capability-Feasibility-Condition2}), there is an upper bound on the operand net firing vectors in Eq. \ref{Eq:FeasibilityMat}.  

It is important to recognize that operand nets may take an arbitrary topology.  Whereas earlier hetero-functional graph theory works on resilience\cite{Farid:2015:ISC-J19,Farid:2014:ISC-C37,Farid:2014:ISC-C38,Thompson:2020:SPG-C68,Thompson:2021:SPG-J46} required operands to have a state evolution described by a single sequence of events (or transitions), the method advanced here allows for operands to have an arbitrary state evolution.  Consequently, operands nets may describe complex operands (i.e. products and services) composed of more basic operands that undergo assembly (i.e. joining) and disassembly (i.e. disjoining).  As the state of such a complex operand evolves, certain transitions of its operand net may be executed simultaneously (i.e. in parallel).  The resulting operand net firing vectors $U_{l_i}[k]$ may each contain more than a single nonzero element at a given moment $k$.  This parallelism is then reflected as multiple nonzero elements in the engineering system net firing vectors $U_S[k]$.  Additionally, it is possible that there are multiple capabilities that realize a given transition in a given operand.  Such a condition is reflected by a multiple nonzero elements in a given row of the operand-capability feasibility matrix $\widetilde{\Lambda_i}$.  Therefore, the plurality of nonzero elements in $U_S[k]$ is driven by the parallelism in $U_{l_i}[k]$ and the multiple feasibility conditions in $\widetilde{\Lambda_i}$.  In such a manner, the calculation of $\widetilde{A}_{Pk}$ enumerates the number of k-capability paths through a hetero-functional graph for an operand net, even when the operand-net itself can take an arbitrary topology and not just that of a single sequence of events.  

To complete the calculation of path enumeration in engineering systems, it is important to recognize that the total number of paths depends on their length; from length 2 up to a prespecified number $K$ of capabilities.  
\begin{align}\label{Eq:AP}
\widetilde{A}_{P} &=\sum_{k=2}^K\widetilde{A}_{Pk}=\sum_{k=2}^K[\widetilde{A}_{Pk-1}]*\Big[\widetilde{A}_{\rho s}\cdot\Big(\mathbf{1}*(U_{S}[k]\cdot\neg{\widetilde{K}}_S^{VT})\Big)\Big]
\end{align}

\vspace{-0.1in}
\subsection{Definition of Performance}\label{Sec:Performance}
In the context of this work, the performance of an engineering system depends purely on its static structure.  Consider an operand $l_i$, that is being delivered by an engineering system through a path that begins with capability $\epsilon_{\psi_1}$ and ends with capability $\epsilon_{\psi_2}$.  If a quantity of this operand $Q_i(\psi_1,\psi_2)$ is delivered between this pair of capabilities $(\epsilon_{\psi_1},\epsilon_{\psi_1})$, then the associated performance of the engineering system is given by\cite{Farid:2015:ISC-J19,Farid:2014:ISC-C37,Farid:2014:ISC-C38,Thompson:2020:SPG-C68,Thompson:2021:SPG-J46}:
\begin{equation}
EP(i,\psi_1,\psi_2)=Q_i \cdot e_{\psi_1}^T\bigg(bi\left[A_{P_i}\right]\bigg)e_{\psi_2}
\end{equation}
Here, it is assumed that the existence of such a delivery path, rather than the number of such delivery paths, is sufficient for the delivery of the entire quantity $Q_i(\psi_1,\psi_2)$ of the operand $l_i$.  As such, it assumes that each path is not capacity limited.  

The above result can be generalized to account for the flow of all operands $l_i \in L$ between any pair of capabilities in the engineering system\cite{Farid:2015:ISC-J19,Farid:2014:ISC-C37,Farid:2014:ISC-C38,Thompson:2020:SPG-C68,Thompson:2021:SPG-J46}.  
\begin{align}\label{Eq:EP2}
EP=\sum_i^{|L|} c_iQ_i\cdot \mathds{1}^T\bigg( bi\left[A_{P_i}\right]\bigg)\mathds{1} 
\end{align}
where $c_i$ is a measure of value of the $i^{th}$ operand such as utility, cost or profit that harmonizes the units of all $Q_i$.  Eq. \ref{Eq:EP2}, when taken in the context of Eq. \ref{Eq:AP}, shows the direct dependence of the performance of an engineering system on its structure as quantified by the choice of operand, its hetero-functional adjacency matrix $A_\rho$ and its system constraints matrix $K_S$.  Consequently, it is useful to write the engineering performance of an engineering systemin terms of these three quantities:  $EP(L,A_\rho,K_S)$.  

\vspace{-0.1in}
\subsection{Actual Engineering Resilience}\label{Sec:AER}
The actual engineering resilience (AER) with respect to a particular disruption that takes the system through the transformation: $(L_0, A_{\rho 0}, K_{S0}) \rightarrow (L, A_{\rho}, K_{S})$ can now be defined straightforwardly\cite{Farid:2015:ISC-J19,Farid:2014:ISC-C37,Farid:2014:ISC-C38,Thompson:2020:SPG-C68,Thompson:2021:SPG-J46}.  
\begin{align}
AER=\frac{EP(L,A_\rho,K_S)}
{EP(L_0,A_{\rho 0},K_{S0})} 
\end{align}
This actual engineering resilience measure benefits from the binary function $bi()$.  As expected, engineering systems that exhibit some path redundancy for their services will not suffer from performance degradation.  Indeed, for many engineering managers, path redundancy may be viewed as unnecessary luxury, or a costly inefficiency, when the sole focus is to deliver products and services through an optimal delivery path.  That said, the $bi()$ also hides the effect of redundancy elimination caused by successive disruptions and so is not the most accurate predictor of the engineering system's ``structural health" or vulnerability towards future disruptions.  

\vspace{-0.1in}
\subsection{Latent Engineering Resilience}\label{Sec:LER}
To address the limitations of the actual engineering resilience measure, a latent engineering resilience measure is proposed\cite{Farid:2015:ISC-J19,Farid:2014:ISC-C37,Farid:2014:ISC-C38,Thompson:2020:SPG-C68,Thompson:2021:SPG-J46}.
\begin{align}
LER=\frac{\displaystyle\sum_i^{|L|} c_iQ_i\cdot \mathds{1}^T\bigg(A_{P_{i}}(K_S)\bigg)\mathds{1}}{\displaystyle\sum_i^{|L_0|} c_iQ_i\cdot \mathds{1}^T\bigg( A_{P_{0i}}(K_{S0})\bigg)\mathds{1}}
\end{align}
where again the number of paths for an operand $l_i$ through hetero-functional graph adjacency matrix $A_{Pi}$ is written explicitly in terms of its dependency on the system constraints matrix $K_S$.  Here, the LER measure degrades gracefully with the transformation $(L_0, A_{\rho 0}, K_{S0}) \rightarrow (L, A_{\rho}, K_{S})$ because of the ratio of actual enumerated paths to the prior enumerated paths.  Furthermore, it is important to recognize that the paths through a hetero-functional adjacency matrix often grows exponentially with number of steps in the path.  Conceptually, this is because the choice of paths through the hetero-functional graph has a corresponding decision-tree with the same number of end-nodes as the number of enumerated paths.  Consequently, it is often useful to report the $LER$ measure as $log(LER)$ \cite{Thompson:2021:SPG-J46} so as to understand how a given disruption can close off one or more large branches of this decision-tree and leave others intact.  

\vspace{-0.1in}
\subsection{Dynamic Engineering Resilience}\label{Sec:DER}
Finally, Fig. \ref{Fig:resilfiig} describes the resilience of an engineering system by its engineering performance as it survives an initial disruption and then bounces back to a new equilibrium level.  In this context, the AER and LER measures provided above are static resilience (i.e. survivability) measures\cite{Farid:2015:ISC-J19,Farid:2014:ISC-C37,Farid:2014:ISC-C38,Thompson:2020:SPG-C68,Thompson:2021:SPG-J46}.  They correspond to the engineering performance of an engineering system at a single point in time; presumably after a disruption.  A dynamic resilience measure must, therefore, recognize that the engineering performance will begin to rise again with restorative actions\cite{Farid:2008:IEM-C05} that evolve $(L[t], A_{\rho}[t], K_{S}[t])$ in time.  Constraints within $K_S[t]$ may be removed, new capabilities in $A_\rho[k]$ may be added, and new operands in the form of products and services in $L[t]$ may be added as well.   Consequently, and much like other literature on dynamic resilience\cite{Sharma:2018:00}, a dynamic measure of actual and latent engineering resilience integrates the static measures over time.  
\begin{align}
DAER &= \frac{1}{T}\sum_{t=0}^T AER[t] \Delta T \\ 
DLER &= \frac{1}{T}\sum_{t=0}^T \log(LER[t]) \Delta T
\end{align}

\liinesbigfig{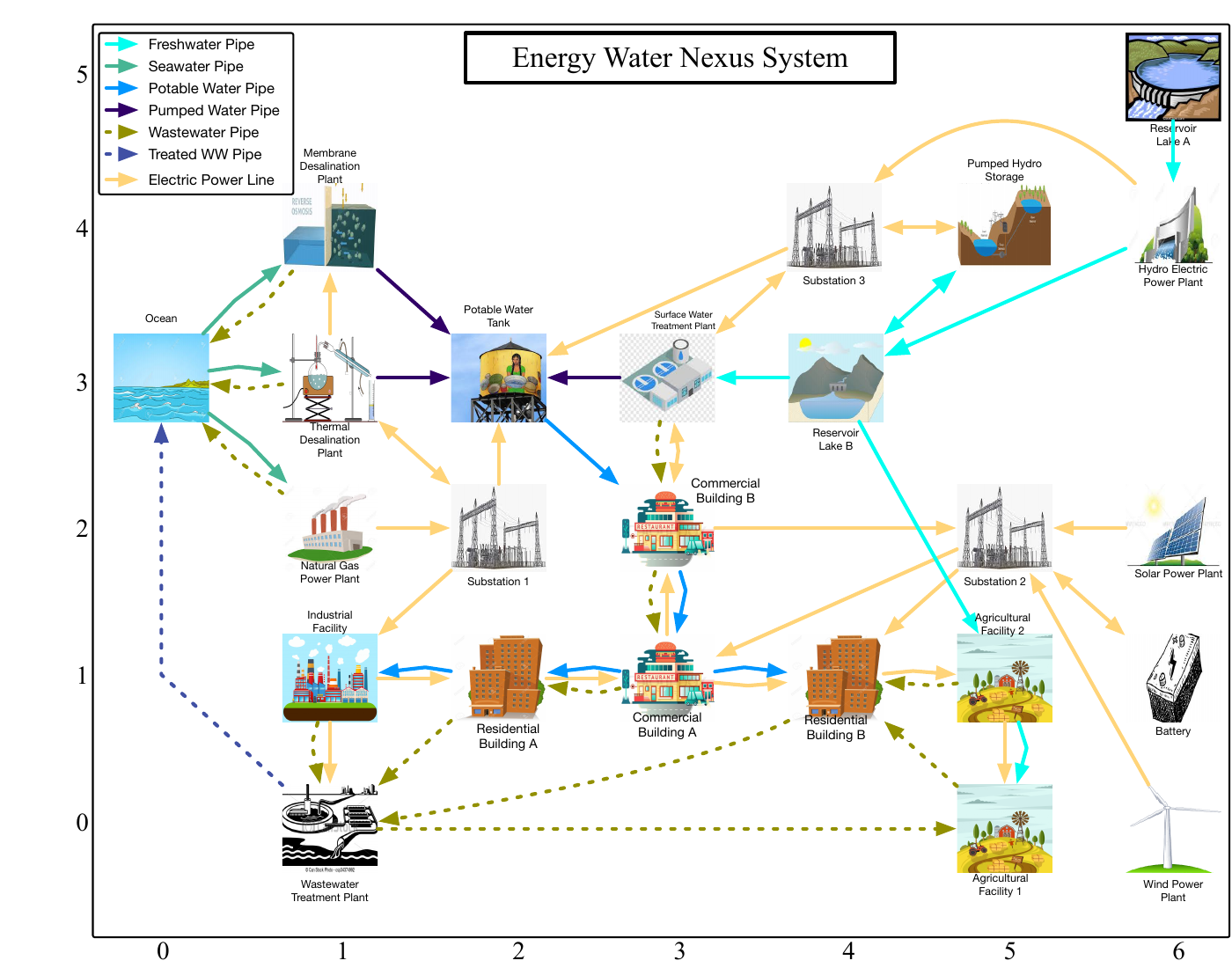}{An Energy Water Nexus Systems is chosen as an  example convergent system-of-systems.  (Adapted from \cite{Thompson:2018:00})}{EWN}

\vspace{-0.1in}  
\section{Numerical Demonstration:  An Example Convergent System-of-Systems}\label{Sec:IllusExample}
Based on the hetero-functional graph theory model and resilience measures developed in Sec. \ref{Sec:HFGT} and \ref{Sec:Resilience} respectively, a hypothetical convergent system-of-systems is chosen for analysis.  Fig. \ref{Fig:EWN} shows a hypothetical Energy Water Nexus (EWN) system as such a convergent system-of-systems.  Importantly, the engineering system has many matter and energy conversion functions that would impede an analysis based upon formal graph theory.  Similarly, its heterogeneity of form and function may exhibit one of the modeling constraints that would impede a multi-layer network analysis.  Furthermore, an energy-water nexus system is by definition a system-of-systems\cite{Lubega:2014:EWN-J11,Lubega:2014:EWN-J12,Lubega:2013:EWN-BC02}.  Consequently, multiple engineering disciplines would, by default, describe their respective portions of this system-of-systems without necessarily converging their perspectives into a single model based upon a common language.  Therefore, this energy-water nexus system is a suitable choice for demonstrating a hetero-functional graph resilience analysis of convergent system-of-systems.    

The convergence of this energy-water nexus system as a system-of-systems is achieved through model-based systems engineering and hetero-functional graph theory.   The convergence process begins with the recognition that the EWN system depicted in Fig. \ref{Fig:EWN} is a partial instance of a recently developed Hydrogen-Energy-Water Reference Architecture (HEWRA)\cite{Farid:2022:ISC-AP83} which was converged from an energy-water nexus reference architecture\cite{Lubega:2014:EWN-J11} and the American Multi-Modal Energy System Reference Architecture\cite{Thompson:2023:ISC-J53}.  While the entire reference architecture cannot be included in this paper, Fig. \ref{Fig:HEWRA} shows the top-level context diagram from which the remainder of the system form, function and concept is elaborated.  It organizes a hydrogen-energy-water system-of-systems into its component systems and perhaps most importantly identifies the exchanges of matter and energy operands between them.  More specifically, the stylized icons in Fig. \ref{Fig:EWN} exist within the electric grid, engineered water supply system, wastewater management system, domestic economy, and natural water system depicted in Fig. \ref{Fig:HEWRA}.  The development of a HEWRA ensures that the entire system-of-systems is converged into a single model stated in SysML\cite{Delligatti:2014:00,Friedenthal:2014:00} as a common language.  

\liinesbigfig{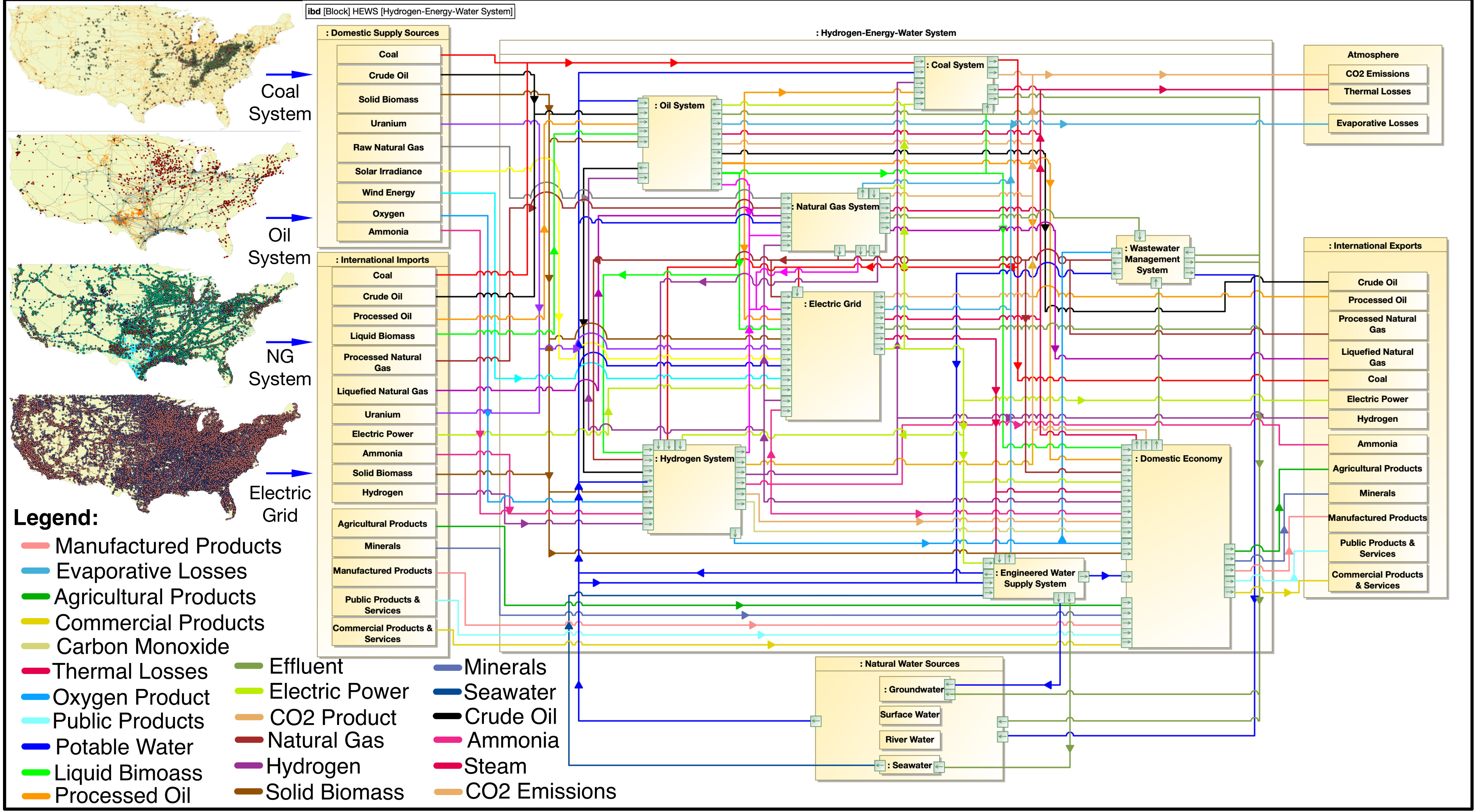}{The top level context diagram of a  Hydrogen-Energy-Water Reference Architecture (HEWRA) \cite{Farid:2022:ISC-AP83,Lubega:2014:EWN-J11,Thompson:2023:ISC-J53})}{HEWRA}

Next, it is important to recognize that the HFGT meta-architecture shown in Fig. \ref{Fig:HFGT-Meta} is a domain-independent generalization of the Hydrogen-Energy-Water Reference Architecture.  Consequently, the entire HEWRA can be restated in terms of hetero-functional graph theory terms.   Moreover, any instance of the HEWRA, like the one depicted in Fig. \ref{Fig:EWN} can also be stated in hetero-functional graph theory terms.  There are 78 system resources (Defn. \ref{Defn:Resource}).  Of these, the 24 stylized icons in Fig. \ref{Fig:EWN} constitute the system buffers.  The 2 reservoir lakes, 1 potable water tank, and 1 Ocean constitute the 4 independent buffers while the other icons represent transformation resources.  Meanwhile, the 23 electric power lines, 3 seawater pipes, six freshwater pipes, 6 potable water pipes, 3 pumped water pipes, 1 treated wastewater pipe, and 12 wastewater pipes constitute the system's 54 transportation resources.  Additionally, adherence to the HEWRA demands that EWN system have 17 operands (Defn. \ref{Defn:Operand}):  seawater, freshwater, potable water, wastewater, treated wastewater, solar irradiance, wind energy, natural gas, electricity, raw materials, agricultural products, commercial products, manufactured products, water vapor loss, process steam, heat loss, and carbon dioxide.  Adherence to the HEWRA also reveals that there are 4047 processes (Defn. \ref{Defn:Process}) that arise from 15 transformation processes, 7 holding processes, and 576 transportation processes.  Of these 4047 processes, only 15 transformation processes and 65 refined transportation processes are used in the instantiated EWN system for a total of 80.  These processes collectively operate on the 17 operands and ensure that all matter and energy conversation laws are respected.  

\liinesbigfig{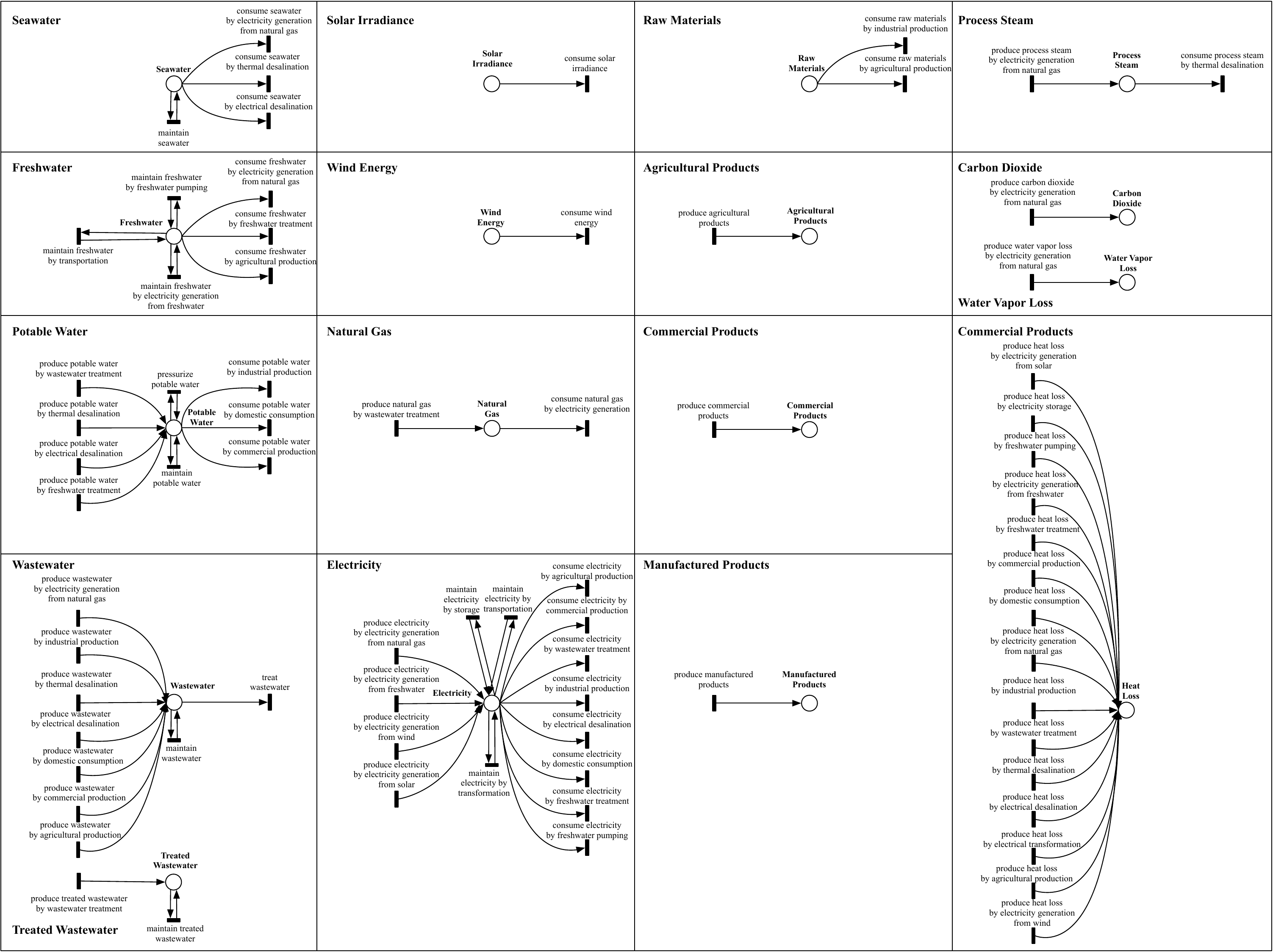}{The 17 Operand Nets of the Energy-Water Nexus System}{OperandNets}

Once the EWN system has been converged into the HFGT meta-architecture, the HFGT model and resilience measures developed in Sec. \ref{Sec:HFGT} and \ref{Sec:Resilience} can be straightforwardly applied.  While it is possible to manually deduce the sets and matrices defined in Defn. \ref{Defn:Capability}-\ref{Defn:SyncMat}, it is prohibitively tedious for all but the simplest of systems.  Instead, a (python-based) HFGT toolbox\cite{Thompson:2023:ISC-JR02} can be used to produce these quantities automatically.  Doing so first requires the preparation of an input XML file that respects the structure and function of the HFGT meta-architecture and HEWRA.  The input XML file associated with the EWN system in Fig. \ref{Fig:EWN} has been included in this paper's supplementary materials.  For this small system, the HFGT toolbox run in approximately 3 seconds on a Macbook Air with an Apple M2 process and 24GB of RAM.  The execution of HFGT toolbox reveals that the system has a total of 86 capabilities (Defn. \ref{Defn:Capability}).  These are organized into a system concept (Defn. \ref{Defn:SystemConcept}) which forms a bipartite graph between the 4047 system processes and 78 system resources.  To represent normal (i.e. undisrupted) operation, the system constraints matrix (Defn. \ref{Defn:SystemConstraint}) is a zero matrix of the same size; creating an equivalence between the system knowledge base (Defn. \ref{Defn:KnowledgeBase}) and system concept.  The HFGT toolbox also straightforwardly calculates the negative and positive $3^{rd}$ hetero-functional incidence tensors (Defn. \ref{Defn:MRhoNeg} and \ref{Defn:MRhoPos} respectively).  As expected, they have a size of 17 operands by 24 buffers by 86 capabilities and a sparsity of approximately 0.3\%.   These tensors are matricized to produce the engineering system net (Defn. \ref{Defn:ESN}), its state transition function (Defn. \ref{Defn:ESN-STF}), and the hetero-functional adjacency matrix (Defn. \ref{Defn:ARho}).  To support the resilience analysis, two more capabilities are added to the analysis so as to represent the beginning and end of the EWN system's operation.  The input XML file to the HFGT toolbox also specifies 17 operands and their respective operand nets (Defn. \ref{Defn:OperandNet}).  For clarity of discussion, each of these have been depicted graphically in Fig. \ref{Fig:OperandNets}.  For each of these operand nets, it also calculates the associated operand-capability feasibility matrix (Defn. \ref{Defn:SyncMat}).  In summary, the HFGT toolbox provides all the prerequisite calculations described in Sec. \ref{Sec:HFGT} so as to support the resilience analysis described in Sec. \ref{Sec:Resilience}.  

One of the strengths of the resilience measures provided in Sec. \ref{Sec:Resilience} is that it recognizes that the resilience of an engineering system depends on the operand being delivered.   A large scale engineering system or system-of-systems may be resilient with respect to the delivery of one operand but not resilient with respect to the delivery another.  For example, a quick inspection of Fig. \ref{Fig:EWN} reveals that there are five types of electric power generation facilities but there is only one wastewater treatment plant.  Consequently, one would expect this EWN system to be more resilient with respect to the delivery of electricity than to the delivery of treated wastewater.  While such an operand-by-operand analysis certainly provides sectoral-level insight, it does not provide a holistic measure of resilience for the EWN nexus as a system-of-systems.  Fortunately, and as Sec. \ref{Sec:OperandBehavior} mentioned, an operand net need not simply represent the state evolution of individual commodities but can also describe the delivery of a complex system-of-systems service $l_{SoS}$ that consists of multiple operands.  To that end, the 17 operand nets shown in Fig. \ref{Fig:OperandNets} can be combined into a single SoS operand net ${\cal N}_{l_{SoS}}= \{S_{l_{SoS}}, {\cal E}_{l_{SoS}}, \textbf{M}_{l_{SoS}}, W_{l_{SoS}}, Q_{l_{SoS}}\}$ that describes the state evolution of all 17 operands together.  This is achieved in five steps steps:
\begin{enumerate}
\item The SoS operand net places are the set union of all of the individual operand net places.  % $S_{l_{SoS}} = \bigcup_i^{|L|} S_{l_{i}}$.  
\item Any transitions in all of the of the individual operand nets that are fired simultaneously are consolidated into a single transition.  For example, the transition ``consume solar irradiance" and the transition ``produce electricity by electricity generation from solar" is replaced with a single transition entitled ``generate electricity from solar irradiance".  
\item Any arcs to/from the simultaneously fired operand net transitions are reconnected to the corresponding consolidated operand net transitions.  
\item An initial transition (with no preset places) is added with arcs to the places associated with seawater, freshwater, solar irradiance and wind energy.  Consequently, all paths begin with the provision of these operands (as natural resources).  
\item An final transition (with no postset places) is added with arcs to the places associated with manufactured products, agricultural products, commercial products, and treated wastewater.  Consequently, all paths end with the creation of these operands as ``must-produce" products.  
\end{enumerate}
The resulting SoS operand net can then be incorporated straightforwardly into the resilience analysis of the EWN system as a system-of-systems.  

The resilience of the EWN system as a system-of-systems is studied for the 30 disruptive scenarios summarized in Table \ref{Ta:Results}.  The first scenario envisions normal operation where the entire EWN system is fully functional.  For simplicity of presentation, the remaining scenarios include disruptions to one of each type of resource.  As an exception, two freshwater pipes and two power lines were selected for disruption scenarios rather than one.  These additional scenarios provide further insight into a resilience analysis founded in hetero-functional graph theory.    Because eigenvalue centrality is often used as a proxy-measure of importance in a network\cite{Barabasi:2016:00}, the scenarios have been ordered from highest to lowest eigenvalue centrality.  In all, the 30 disruption scenarios include the failure of transformation resources, independent buffers, and transportation resources.  It is also important to recognize that such an eigenvalue centrality analysis would have been impossible on formal graphs and multi-layer networks because eigenvalue centrality analysis only pertains to nodes and not edges.  In contrast, because hetero-functional graphs use capabilities as nodes, they place point facilities (e.g. buildings) and line facilities (e.g. water pipes) on an equal analytical footing.  In fact, Table \ref{Ta:Results} shows that the eigenvalue centrality of the Freshwwater Pipe 2 capability is equivalent to that of the Reservoir Lake B capability.  Finally, the resilience analysis is conducted on the basis of paths consisting of up to 10 capabilities and the dynamic resilience measures are calculated assuming that each disruption scenario has the same duration.  

\vspace{-0.1in}
\begin{longtable}{p{0.45in}p{2.0in}p{0.75in}rll}
\caption{Resilience Analysis of the EWN Systesm} \label{Ta:Results} \vspace{-0.1in}\\ \toprule 
\textbf{Scenario \newline Index} & \textbf{Disruption \newline Scenario}	& \textbf{Eigenvalue \newline Centrality}	& \textbf{Paths}  & \textbf{AER} & \textbf{LER}\\\midrule
1  &Normal Operation & NaN & 1840 & 1 & 1.0\\
2  &Pumped Hydro Storage & 0.41764 & 833 & 1 & 0.45272\\
3. &Freshwater Pipe 3 & 0.3995 & 1213 & 1 & 0.65924\\
4. &Freshwater Pipe 2 & 0.16285 & 1757 & 1 & 0.95489\\
5. &Reservoir Lake B & 0.16285 & 1567 & 1 & 0.85163\\
6. &Power Line 15 & 0.13384 & 1297 & 1 & 0.70489\\
7. &Power Line 1 & 0.13384 & 560 & 1 & 0.30435\\
8. &Hydro Electric Power Plant & 0.09193 & 1710 & 1 & 0.92935\\
%&Freshwater Pipe 1 & 0.07379 & 1803 & 1 & 0.97989\\
%&Power Line 23 & 0.06267 & 1774 & 1 & 0.96413\\
%&Power Line 2 & 0.06267 & 1454 & 1 & 0.79022\\
%&Power Line 13 & 0.06267 & 1357 & 1 & 0.7375\\
9. &Substation 3 & 0.06267 & 1235 & 1 & 0.6712\\
10 &Reservoir Lake A & 0.03008 & 1803 & 1 & 0.97989\\
%&Power Line 12 & 0.01989 & 1368 & 1 & 0.74348\\
%&Power Line 10 & 0.00601 & 1744 & 1 & 0.94783\\
%&Power Line 6 & 0.00174 & 1763 & 1 & 0.95815\\
%&Power Line 18 & 0.00174 & 1698 & 1 & 0.92283\\
%&Power Line 21 & 0.00085 & 1756 & 1 & 0.95435\\
%&Power Line 22 & 0.00085 & 1756 & 1 & 0.95435\\
%&Power Line 20 & 0.00085 & 1692 & 1 & 0.91957\\
%&Substation 2 & 0.00085 & 1574 & 1 & 0.85543\\
%&Power Line 11 & 0.00085 & 1474 & 1 & 0.80109\\
11 &Battery & 0.00035 & 1677 & 1 & 0.91141\\
12 &Solar Power Plant & 0.00025 & 845 & 1 & 0.45924\\
13 &Wind Power Plant & 0.00025 & 842 & 1 & 0.45761\\
%&Power Line 8 & 0.00015 & 1737 & 1 & 0.94402\\
%&Substation 1 & 7.0e-5 & 1840 & 1 & 1.0\\
%&Power Line 3 & 7.0e-5 & 1789 & 1 & 0.97228\\
%&Power Line 4 & 7.0e-5 & 1743 & 1 & 0.94728\\
14 &Natural Gas Power Plant & 2.0e-5 & 1298 & 1 & 0.70543\\
15 &Seawater Pipe 3 & 1.0e-5 & 1833 & 1 & 0.9962\\
16 &Residential Building A & 0.0 & 1840 & 1 & 1.0\\
%&Power Line 7 & 0.0 & 1840 & 1 & 1.0\\
%&Power Line 16 & 0.0 & 1840 & 1 & 1.0\\
%&Power Line 17 & 0.0 & 1840 & 1 & 1.0\\
%&Seawater Pipe 1 & 0.0 & 1840 & 1 & 1.0\\
%&Freshwater Pipe 4 & 0.0 & 1840 & 1 & 1.0\\
17 &Potable Water Pipe 6 & 0.0 & 1840 & 1 & 1.0\\
18 &Wastewater Pipe 2 & 0.0 & 1840 & 1 & 1.0\\
%&Wastewater Pipe 3 & 0.0 & 1840 & 1 & 1.0\\
%&Wastewater Pipe 4 & 0.0 & 1840 & 1 & 1.0\\
%&Wastewater Pipe 8 & 0.0 & 1840 & 1 & 1.0\\
%&Wastewater Pipe 9 & 0.0 & 1840 & 1 & 1.0\\
%&Wastewater Pipe 11 & 0.0 & 1840 & 1 & 1.0\\
%&Seawater Pipe 2 & 0.0 & 1839 & 1 & 0.99946\\
%&Freshwater Pipe 6 & 0.0 & 1839 & 1 & 0.99946\\
%&Wastewater Pipe 12 & 0.0 & 1835 & 1 & 0.99728\\
%&Power Line 5 & 0.0 & 1834 & 1 & 0.99674\\
19 &Pumped Water Pipe 1 & 0.0 & 1834 & 1 & 0.99674\\
20 &Ocean & 0.0 & 1832 & 1 & 0.99565\\
21 &Surface Water Treatment Plant & 0.0 & 1829 & 1 & 0.99402\\
22 &Commercial Building B & 0.0 & 1819 & 1 & 0.98859\\
% &Wastewater Pipe 7 & 0.0 & 1819 & 1 & 0.98859\\
%&Potable Water Pipe 4 & 0.0 & 1818 & 1 & 0.98804\\
%&Potable Water Pipe 5 & 0.0 & 1818 & 1 & 0.98804\\
%&Commercial Building A & 0.0 & 1815 & 1 & 0.98641\\
%&Pumped Water Pipe 2 & 0.0 & 1808 & 1 & 0.98261\\
23 &Potable Water Tank & 0.0 & 1807 & 1 & 0.98207\\
24 &Wastewater Pipe 5 & 0.0 & 1794 & 1 & 0.975\\
%&Wastewater Pipe 6 & 0.0 & 1794 & 1 & 0.975\\
%&Pumped Water Pipe 3 & 0.0 & 1793 & 1 & 0.97446\\
%&Wastewater Pipe 1 & 0.0 & 1793 & 1 & 0.97446\\
%&Wastewater Pipe 10 & 0.0 & 1793 & 1 & 0.97446\\
%&Potable Water Pipe 3 & 0.0 & 1777 & 1 & 0.96576\\
%&Power Line 14 & 0.0 & 1760 & 1 & 0.95652\\
%&Potable Water Pipe 2 & 0.0 & 1748 & 1 & 0.95\\
%&Potable Water Pipe 1 & 0.0 & 1741 & 1 & 0.9462\\
%&Power Line 9 & 0.0 & 1733 & 1 & 0.94185\\
%&Power Line 19 & 0.0 & 1631 & 1 & 0.88641\\
% &Residential Building B & 0.0 & 1541 & 1 & 0.8375\\
25 &Membrane Desalination Plant & 0.0 & 1478 & 1 & 0.80326\\
26 &Agricultural Facility 1 & 0.0 & 1468 & 1 & 0.79783\\
%  &Freshwater Pipe 5 & 0.0 & 1333 & 1 & 0.72446\\
% &Agricultural Facility 2 & 0.0 & 1253 & 1 & 0.68098\\
27 &Treated Wastewater Pipe & 0.0 & 1192 & 1 & 0.64783\\
28 &Thermal Desalination Plant & 0.0 & 833 & 1 & 0.45272\\
29 &Wastewater Treatment Plant & 0.0 & 0 & 0 & 0.0\\
30 &Industrial Facility & 0.0 & 0 & 0 & 0.0\\
 -- & Dynamic Resilience 			& -- 			& --		& $\frac{28}{30}$ & 0.87062 			\\\bottomrule        
 \end{longtable}

Table \ref{Ta:Results} summarizes the results of the resilience analysis.  In normal operation, the EWN system has 1840 paths for delivering the system-of-systems service consisting of all 17 operands.  As this scenario is the base case, it is associated with actual and latent engineering resilience values of one.  The analysis for the remaining scenarios provides interesting insights.   It becomes immediately clear that there is no correlation between the number of paths and the eigenvalue centrality of a disrupted resource.  While the disruption of the highly central pumped hydrostorage removes \~55\% of the paths, the disruption of the significantly less central power line removes \~70\% of the paths.   This numerical result reconfirms the results of a similar analysis conducted on the entire American electric power system\cite{Thompson:2021:SPG-J46}.  Consequently, eigenvalue centrality is best restricted to \emph{homo-functional} engineering systems where formal graphs can be used for analysis.  In contrast, the hetero-functionality found in this EWN system (and its HEWRA) imposes a certain flow of functions that determines whether some capabilities will be critical, redundant or entirely superfluous to the path of delivery.  For example, although the wastewater treatment plant has a zero eigenvalue centrality, its failure eliminates \emph{all} 1840 delivery paths because an EWN system \emph{must} treat wastewater in order for it to reliably function.  Similarly, the failure of the pumped hydro storage, hydro-electric power plant, solar power plant, and wind power plants led to strong reductions in the number delivery paths even though they have dramatically different eigenvalue centrality values.  These strong reductions in the number of delivery paths is explained by the proximity of these resources to the natural environment (i.e. system boundary) and the operands that the natural environment provides.  Ultimately, because the system-of-system operand net begins and ends with the natural environment (as explained above), the delivery paths do as well.  Consequently, the disruption of any one of these facilities at the system boundary eliminates between 35-55\% of the delivery paths.  In contrast, the disruption of Residential Building A has no effect on the delivery paths because its capabilities (as defined in the HEWRA) do not contribute to the path of delivery of agricultural products, manufactured products, commercial products, and treated wastewater.   Furthermore, potable water pipe 6 and wastewater pipe 2 have no impact on the delivery of these critical services.  Meanwhile, the failure of the only industrial facility means that the EWN systems will not be able to produce any manufactured products.  

Table \ref{Ta:Results} also highlights the importance of and relationship between enumerated delivery paths, actual engineering resilience, latent engineering resilience and dynamic engineering resilience as quantitative measures.  Quite interestingly, the EWN system is highly resilient.  The wastewater treatment plant and the industrial facility are the only two single points of failure.  Under all other disruption scenarios, the EWN system continues to provide a multiplicity of delivery paths and would continue to function accordingly.  For this reason, and quite practically, the actual engineering resilience retains a perfect value of one for all but two disruption scenarios.  Despite these positive results in the actual engineering resilience, they can perhaps give a false sense of security because the AER measure hides the reduction of enumerated delivery paths caused by subsequent disruptions until the very last delivery path is lost -- potentially abruptly.  In contrast, the latent engineering resilience measure degrades gracefully showing reductions of enumerated delivery paths with each passing disruption.  It reveals that when certain disruptions occur to critical, redundant, or superfluous resources, they have highly disparate impacts on the number of delivery paths.  In that regard, it provides a quantitative measure of a system's ``structural health" even when no single disruption is sufficient to stop the delivery of a complex service.  

\vspace{-0.1in}
\section{Conclusion}\label{Sec:Conclusion}
This paper develops a methodology for hetero-functional graph resilience analysis and demonstrates it on a convergent system-of-systems.  In that regard, it addresses the recognized need for resilience in systems-of-systems.  Furthermore, the reliance on the Systems Modeling Language (SysML), Model-based Systems Engineering (MBSE), and Hetero-functional graph Theory (HFGT) provides a means of convergence of these systems-of-systems based upon reconciled ontologies, data, and theoretical methods.  The paper includes both the ``survival" as well as ``recovery" components of resilience.  It also strikes a middle ground between two disparate approaches to resilience measurement; namely structural measurement of formal graphs and detailed behavioral simulation.  This paper also generalizes a previous resilience measure based on HFGT.  More specifically, the actual, latent, and dynamic engineering resilience measures now accommodate systems-of-systems that deliver complex services of arbitrary behavioral topology.  The developed resilience measures also benefit from tensor-based theoretical developments and toolbox-based computational developments in HFGT.  Finally, a resilience analysis is conducted on a hypothetical energy-water nexus of moderate size as a type of systems-of-systems.  The resilience measure is able to identify and differentiate system capabilities that are critical, redundant, and superfluous to system-of-systems service delivery.  It is able to differentiate between the actual versus latent engineering resilience of an engineering system where the formerer addresses actual service delivery and the latter addresses ``structural health"  or vulnerability to future disruptions.  Finally, this hetero-functional graph analysis shows that resilience in system-of-systems does not correlate with eigenvalue-centrality as it is commonly understood it does in homo-functional network systems.  
  
The original contribution presented in this paper leaves open many opportunities for future work.  As highlighted in Sec. \ref{Sec:FormalGraphTheory} and \ref{Sec:Multi-LayerNetworks}, hetero-functional graph theory is providing, for the first time, an ability to study the structure and function of systems-of-systems of arbitrary topology.  Consequently, the resilience measures developed here may be applied to many systems-of-systems application domains like electrified transportation\cite{Farid:2016:ETS-J27,vanderWardt:2017:ETS-J33} and multi-energy system\cite{Thompson:2024:ISC-J55}.  Second, some disruptions may not be binary either in the sense that they may lead to impaired rather than fully disabled functionality or that they have a fundamentally probabilistic behavior.  Finally, the authors believe that this work has direct application to resilient control, decision-making, and management systems.

\vspace{-0.15in}
\bibliographystyle{IEEETran}
\bibliography{LIINESLibrary,LIINESPublications}
\vspace{-0.45in}
\begin{IEEEbiography}[{\includegraphics[width=1in,height=1.25in,clip,keepaspectratio]{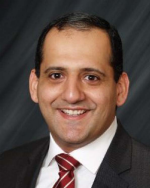}}]{Amro M. Farid}
is the Alexander Crombie Humphreys Chair Professor in Economics of Engineering at the School of Systems and Enterprises at the Stevens Institute of Technology.  He is also the Principal Systems Scientist for the Smart Energy Mission at CSIRO ? Australia?s National Science Agency.  He is also the founding CEO of Engineering Systems Analytics LLC.  Prof. Farid received his Sc.B and Sc.M degrees from MIT and completed his Ph.D. degree at the Institute for Manufacturing within the University of Cambridge Engineering Department in 2007. He leads the Laboratory for Intelligent Integrated Networks of Engineering Systems (LIINES) and has has authored over 160 peer-reviewed publications and 140 invited presentations in Smart Power Grids, Hydrogen-Energy-Water Nexus, Electrified Transportation Systems, Industrial Production \& Supply Chain Energy Management, Smart Cities, Regions \& Nations.  In 2021, he became a Fulbright Future Scholar to investigate the energy-water-hydrogen nexus in Australia. 
\end{IEEEbiography}
\end{document}